\documentstyle[aps,epsf,preprint]{revtex}

\begin{document}

\baselineskip=24 pt


\title
{Monte Carlo simulation of an strongly coupled XY model in three
dimensions}
\author
{ Rasool Ghanbari $^a$ , Farhad Shahbazi $^b$   }
\address
{ \it $^a$ Dept. of Physics , Islamic Azad University, Majlesi
branch, 86315/111, Isfahan ,Iran.\\
\it $^b$ Dept. of Physics , Isfahan University of Technology,
84156, Isfahan, Iran.\\}

\maketitle

\begin{abstract}
Many experimental studies, over the past two decades, have
constantly reported  a novel critical behavior for the transition
from Smectic-A phase of liquid crystals to Hexatic-B phase with
non-XY critical exponents. However according to symmetry
arguments this transition must belong to XY universality class.
Using an optimized  Monte Carlo simulation technique based on
multi-histogram method, we have investigated phase diagram of a
coupled XY model, proposed by Bruinsma and Aeppli (PRL {\bf 48},
1625 (1982)), in three dimensions. The simulation results
demonstrate the existence of a tricritical point for this model,
in which two different orderings are established simultaneously.
This result verifies the accepted idea  the large specific heat
anomaly exponent observed for SmA-HexB transition could be  due to
the occurrence of this transition in the vicinity of a tricritical
point.
\end{abstract}
\hspace{.3in}
\newpage

\section{Introduction}
According to Kosterlitz, Thouless, Halperin, Nelson and Young
(KTHNY) theory {\cite{KT},\cite{hn},\cite{y}}, two dimensional
systems during melting transition from solid to isotropic liquid
go through an intermediate phase called hexatic phase for systems
that have six-fold(hexagonal) symmetry in their crystalline
ground state. This hexatic phase displays short range positional
order, but quasi long range bond-orientational order, which is
different from the true long range bond-orientational and quasi
long range positional order in 2D solid phases. It is known that
for two dimensional systems, the transition from the isotropic
liquid  to hexatic phase could be either a KT transition or a
first order transition {\cite{5}}.

The idea of hexatic phase was first applied to three dimensional
systems by Birgeneau and Lister, who showed that some
experimentally observed smectic liquid crystal phases ,consisting
of stacked 2D layers, could be physical realization of 3D
hexatics\cite{birg}. Assuming that the weak interaction between
smectic layers could make the quasi long range order of two
dimensional layers truly long ranged, they suggest that the 3D
hexatic phases in highly anisotopic systems, possess  short range
positional and true long range bond-orientational order.

The first signs for the existence of the hexatic phase in three
dimensional systems were observed in x-ray diffraction study of
liquid crystal compound
65OBC(n-alkyl-4-m-alkoxybiphenyl-4-carboxylate,n=6,m=5)\cite{huang,pindak},
where a hexagonal pattern of diffuse spots was found in intensity
of scattered x-rays. In addition to this hexagonal pattern, it was
also found that some broader peaks were appeared in the diffracted
intensity which indicate the onset of another ordering. These
broad peaks are related to packing of molecules according to the
herringbone structure perpendicular to the smectic layer stacking
direction. The accompanying of the long range hexatic and short
range herringbone orders make this phase a physically rich phase
which simply is called Hexatic-B (HexB) phase. When temperature is
decreased, the HexB phase transforms via a first order phase
transition into the crystal-E (CryE) phase, which exhibits both
long range positional and long range herringbone orientational
orders. Subsequently, it was found that other components in nmOBC
homologous series (like 37OBC and 75OBC) and a number of binary
mixtures of n-alkyl-4'-n-decycloxybiphenyl-4-carboxilate
(n(10)OBC) with n ranging from 1 to 3 and also the compound
4-propionyl-4'-n-heptanoyloxyazo-benzene (PHOAB) represent
smA-HexB transition, which for later the transition has found to
be clearly first order.

Due to the sixfold symmetry of hexatic phase, the corresponding
order parameter is defined by
$\Psi_{6}=|\Psi_{6}|\exp(i6\psi_{6})$. The U(1) symmetry of the
$\Psi_{6}$, implies that SmA-HexB transition be a member of XY
universality class. However, heat capacity measurements on bulk
samples of 65OBC \cite{huang,huang2} and other calorimetric
studies on many other components in the nmOBC homologous series
\cite{huang,pitch} have yielded very sharp specific heat anomalies
near SmA-HexB transition with no detectable thermal hystersis and
with very large value for the heat capacity critical exponent,
$\alpha\approx0.6$. These results indicate that this is a
continues (second order) phase transition, but not belonging to
The 3D XY universality class, for which the specific heat critical
exponent is nearly zero ($\alpha\approx-0.007$\cite{zinn}).On the
other hand, the other static critical exponents determined  from
thermal conductivity ($\eta=-0.19$)and  birefringence experiments
($\beta=0.19$) \cite{huang}, all differ from the 3D XY values,
indicating a novel phase transition with probably a new
universality class.

It is also interesting to mention that the same  heat capacity
measurement studies of (truly two-dimensional) two-layer free
standing films of different nmOBC compound result a second order
SmA-HexB transition, described by the heat capacity exponent
$\alpha\approx 0.3$\cite{huang,huang3}. This is obviously  in
contrast with the usual broad and nonsingular specific heat hump
of the KT transition in the 2D XY model,  suggesting that SmA-HexB
transition can not be described  simply  by a unique  XY order
parameter.

The unusual aspects of SmA-HexB transition  in two and three
dimensions, have attracted the interests of physicists in the past
two decades. The first theoretical attack to this problem was
done by Bruinsma and Aeppli{\cite{BA}} who formulated a
Ginzburg-Landau theory that included both hexatic and herringbone
order. Because of the broadness of x-ray diffracted peaks
associated to herringbone order (which is the reason of being
short rang), they considered an XY order parameter with two fold
symmetry for herringbone ordering
$(\Phi_{2}=|\Phi_{2}|\exp(i2\phi_{2}))$ and also based on
symmetry arguments, they made a minimal coupling between the
hexatic and herringbone order parameters as
$V_{hex-her}=hRe(\Psi_{6}^{*}\Phi_{2}^{3})$. Microscopically, the
origin of this coupling could be the  anisotropy presented in
liquid crystals molecular structures\cite{michel1,michel2}.

In the mean field approach their results indicate that the
SmA-HexB transition should be continuous. However one-loop
renormalization calculations show that short range molecular
herringbone correlations coupled to the hexatic ordering drive
this transition first order, which becomes second order at a
tricritical point{\cite{BA}}. Their result indicates the
existence of two tricritical points, one for the transition
between SmA phase ($\Psi=0, \Phi=0$) and the stacked hexatic
phase ($\Psi\neq 0, \Phi=0$), and another for the transition
between the SmA and the phase possessing both hexatic and
herringbone order ($\Psi\neq 0, \Phi\neq 0$). Therefore, They
concluded that the occurrence of phase transition  near the
tricritical points, with heat capacity exponent $\alpha=0.5$,
would be a good explanation for large heat capacity exponents
observed in the  experiments. Recently, the RG calculation of BA
model has been revised in \cite{kohan} which resulted in finding
another non-trivial fixed point missed in original work of
Bruinsma and Aeppli. But it has been shown that this new fixed
point is unstable in one loop level (order of $\epsilon$), which
refuses this fixed point to represent a novel phase transition.
Improvement of this calculation to two loop level (order of
$\epsilon^{2}$), although make this new fixed point stable, but
gives a  small and negative value for the corresponding heat
capacity anomaly exponent \cite{fash}, which indicts that this
critical point  can not explain the experimental results.
However, the limitations of  RG methods which mostly rely on
perturbation expansions, make them insufficient for accessing  the
strong coupling regimes where one expect that some kind of new
treatment to appear. For this purpose, the numerical simulations
would be useful.

The first  numerical simulations for investigating the nature of
the SmA-HexB transition in 2D systems have been done by Jiang et
al who have used a model consists of a 2D lattice of coupled XY
spins based on the BA Hamiltonian in strong coupling
limit\cite{mc1,mc2}. Their simulation results suggest the
existence of a new type phase transition in which two different
orderings are simultaneously established through a continuous
transition with heat capacity exponent $\alpha\sim 0.3$, in  good
agreement with experimental values.

The success of BA model in two dimensions and also the  absence of
any numerical simulation in three dimensions were our motivations
to investigate numerically the 3-dimensional BA model in strong
coupling limit .To do this, we  employ a  high resolution Monte
Carlo simulation based on multi-histogram method.

The rest of this paper is organized as follows. In section. II, we
introduce model Hamiltonian  and give a brief introduction to
optimized  Monte Carlo method based on multiple histograms and
also Some methods for  analyzing the  Monte Carlo data, to
determine the order of transitions. The simulation results and
discussion is given in section III and conclusions will appear in
section IV.

\section{Monte Carlo simulation}

\subsection{Model Hamiltonian}

Recalling  the six-fold symmetry of hexatic order and  two-fold
symmetry of the herringbone order, the Hamiltonian which describes
both orderings ought to be invariant with respect to the
transformation $\Phi\rightarrow\Phi+n\pi$ and
$\Psi\rightarrow\Psi+m(2\pi/6)$ where $n$ and $m$ are integers.
Thus to lowest order in $\Psi$ and $\Phi$, one can write the
following  Hamiltonian for BA model:

\begin{eqnarray}
H=&-&J_{1}\sum_{<ij>} \cos(\Psi_{i}-\Psi_{j})- J_{2}\sum_{<ij>}
\cos(\Phi_{i}-\Phi_{j})\nonumber\\
&-&J_{3}\sum_{i} \cos(\Psi_{i}-3\Phi_{i}),
\end{eqnarray}

where the coefficients $J_1$ and $J_2$ are the nearest-neighbor
coupling constants for the bond-orientational order $(\Psi)$ and
herringbone order $(\Phi)$, respectively. The coefficient $J_3$
denotes the coupling strength between these two types of order at
the same 3D lattice site. we are interested in situations in
which $\Psi$ and $\Phi$ are coupled strongly. Therefore we fixed
$J_3=3.0$,larger that both $J_1$ and $J_2$ for all the
simulations. Let assume $J_{1}>J_{2}$, so for sufficiently high
temperatures,(say $T>J_{3}$), the system is in completely
disordered phase. For $T_{c1}<T<J_3$, the system remains
disordered but the phases of the two order parameters become
coupled through the herringbone-hexatic coupling term $J_{3}$. In
mean field level, for $T_{c2}< T <T_{c1}$, bond orientational
order is established through a continuous XY transition and the
ordered state corresponds to $\Psi_{i}\approx\Psi_{j}$ for all
sites i and j, producing  three degenerate minima for the free
energy. So for these range of temperatures the BA Hamiltonian
describes a system with the symmetry of the three-state potts
model and since the ordering transition for three-state potts
model is first order in 3D, the transition between the pure
hexatic and hexatic plus herringbone phases ($\Psi\neq 0,
\Phi\neq 0$) should be first order at $T_{c2}$. Thus for
$J_{2}<J_{1}<J_{3}$ the model exhibits an XY transition at
$T_{c1}$ and a three-state potts-like transition upon decreasing
the temperature down to $T_{c2}${\cite{Jiang}}. For $J_{2} >
J_{1}$, the herringbone order would establish first and cause the
correspondent field $\Phi$ to take nearly the same value for all
sites. Because of this, the coupling term $J_{3}$ acts like a
field on $\Psi$ and so the hexatic order parameter takes a
nonzero value.

The above discussions results that the phase diagram of the BA
model, in mean field level, consists of  three phase transition:
1)A second order Transition from disordered to hexatic phase, 2)A
second order transition from disordered to locked phase consist
of hexatic plus herringbone orders and 3)A first order transition
from hexatic to hexatic plus herringbone phases.

To obtain a qualitative picture of transitions and also the
approximate location of the critical points, we first set a low
resolution simulations. The Simulations were carried out using
standard Metropolis spin-flipping algorithm with six lattice
sizes (L=6,7,8,,9,10,12). During each simulation step, The angles
$\Psi_{i}$ and $\Phi_{i}$ were treated as unconstrained,
continuous variables. The random-angles rotations
($\Delta\Psi_{i}$ and $\Delta\Phi_{i}$) were adjusted  in such a
way that roughly $50\%$ of the attempted angle rotations were
accepted. To ensure thermal equilibrium, 100 000 Monte Carlo
steps (MCS) per spin were used for each temperature and 200 000
MCS were used for data collection.

We have obtained the heat-capacity data as a function of
temperature, shown in Fig. (1) for $J_{1}=1.0$ and $J_{2}=0.5$
and for $J_{2}=0.7,0.8,0.9,1.2$ in Fig. (2).
 Near the lower temperature transition point ($1.2<T<1.35$)
the calculated data were obtained by optimized reweighting using 5
histograms near $T=1.25$(section III). From the preceding
discussion, it is clear that the small broad peak near T=2.2
signals the XY transition due to the $J_{1}$ term, while the
sharp peak located at $T\sim1.25$ is expected to signal a
transition into the state of three-state potts symmetry. The same
simulations based on single spin flipping  algorithm whose
results are represented in Fig.(2), show that the first peak (XY
transition) would disappear for $J_2>0.9$ and therefore only one
transition occurs for those values of $J_{2}$, which verifies
that for these values of $J_{2}$, the transition from disordered
to  herringbone phase, simultaneously  induces hexatic ordering .

To determine The location of the transition temperatures and other
thermodynamic quantities such as specific heat near the
transition points we need to use high resolution methods. For
this purpose we used  multiple-histogram reweighting method
proposed by Ferrenberg and Swendsen {\cite{fs}}, which makes it
possible to obtain accurate data over the transition region from
just a few Monte Carlo simulations.

\subsection{Histogram Method}
 The central idea behind the histogram method is to build up
information on the energy probability density function
$P_{\beta}(E)$, where $\beta=1/T$ is inverse temperature (in units
with $k_{B}=1$). A histogram $H_{\beta}(E)$ which is the number of
spin configurations generated between $E$ and $E+\delta E$.
$P_{\beta}(E)$ is  defined as :

\begin{equation}
 P_{\beta}(E_{i})=\frac{H_{\beta}(E_{i})}{Z_{\beta}},
\end{equation}

where

\begin{equation}
 Z_{\beta}=\sum_{i} H_{\beta}(E_{i}).
\end{equation}

On the other hand we now that $P_{\beta}(E_i)$ is proportional to
the Boltzmann weight $\exp(-\beta E_{i})$ as:

\begin{equation}
P_{\beta}(E_{i})=\frac{g(E_{i})\exp(-\beta E_{i})}{Z_{\beta}},
\end{equation}

in which $g(E_{i})$ is the density of states with energy $E_{i}$
and is independent of temperature. By knowing the probability
distributions in a specific temperature, we can derive the
density of states and find the probability distribution of energy
at any temperature $\beta^{'}$ as follows:

\begin{equation}
P_{\beta^{'}}(E_{i})=\frac{P_{\beta}(E_{i})\exp[(\beta-\beta^{'})E_{i}]}
{\sum_{j}P_{\beta}(E_{j})\exp[(\beta-\beta^{'})E_{j}]}.
\end{equation}

In principle, $P_{\beta}(E)$ only provides information on the
energy distribution of nearby temperatures. This is because the
counting statistics in the wings of the distribution
$H_{\beta}(E)$, far from the average energy at temperature $T$,
will be poor.

To improve the estimation for density of states, one can  take
data at more than one temperature and combine the resultant
histograms so as to take the advantages of the regions where each
provide the best estimate for the density of states. This method
has been studied by Ferrenberg and Swendsen who presented an
efficient way for combining the histograms {\cite{fs}}. Their
approach relies on first determining the characteristic
relaxation time $\tau_{j}$ for the $j$th simulation and using
this to produce a weighting factor $g_{j}=1+2\tau_{j}$. The
overall probability distribution at coupling $K=\beta J$ obtained
from $n$ independent simulation, each with $N_{j}$
configurations, is then given by :

\begin{equation}\label{mh}
P_{K}(E)=\frac{[\sum_{j=1}^{n}g_{j}^{-1}H_{j}(E)]e^{-KE}}
{\sum_{j=1}^{n}N_{j}g_{j}^{-1}e^{-K_{j}E-f_{j}}},
\end{equation}

where $H_{j}(E)$ is the histogram for the $j$th simulation and
the factors $f_{j}$ are chosen self-consistently using
Eq.(\ref{mh}) and

\begin{equation}
e^{f_{j}}=\sum_{E}P_{K_{j}}(E).
\end{equation}

Thermodynamic properties are determined, as before, using this
probability distribution, but now the results would be valid over
a much wider range of temperatures than for any single histogram.
In addition, this method gives an expression for the statistical
error of $P_{K}(E)$ as:

\begin{equation}
\delta P_{K}(E)=[\sum_{j=1}^{n}g_{j}^{-1}H_{j}(E)]^{-1/2}P_{K}(E),
\end{equation}

from which it is clear that the statistical error will be reduced
when more MC simulations are added to the analysis.

\subsection{Order of the transition}

One of the main problems in Monte Carlo data analysis of phase
transitions is determining the order of the transition. Strong
first-order transitions will show marked discontinuities in
thermodynamic quantities such as internal energy and the order
parameter and present no real problems. Weakly first-order
transition are much more difficult to recognize. To understand
the situation, consider a first order phase transition in an
infinitely extended system, for which the correlation length
reaches a finite value $\xi_{c}$ at the transition point where
the phase of the system changes discontinuesly. If $\xi_{c}$ is
too large ,i.e $\xi_{c}>>L$ where $L$ is the linear size of the
system on which the simulation is being done, then the system
would appear to be in the critical region of a continues
transition and it would be very difficult to detect the
discontinuities. However, during the past decades, There have
been significant advances in overcoming this problem. Below we
list a number of techniques for detecting a first-order
transition:\newline
(1) Discontinuities in the internal energy and the order parameter.\\
(2) Hysteresis in the internal energy and the order parameter.\\
(3) Double peaks in the probability density function $P(E)$.\\
(4) The divergence of specific heat  as
$L^{d}$, where $d$ is the spatial dimension.\\
(5) Decreasing the Half-width of the specific heat peak like $L^{-d}$.\\
(6)  The size dependence of the minima of Binder fourth energy
cumulant

\begin{equation}
 U_{4}(L)=1-\frac{<E^4>}{3<E^2>},
\end{equation}

whose value approaches $2/3$ for a continuous transition and some
nontrivial value $U^{*}<{2\over 3}$ at a first-order transition.

The first method as previously mentioned, is inefficient for
weakly first order transitions. The second and third Methods are
based on the fact that the state of a given system representing
first order transition, during its evolution,  may trap, for a
relatively long time, in some local minima  of free energy (called
meta-stable states). these two methods are also  unreliable
because  if the free-energy barrier  is small enough, both phases
will be sampled within time scale of the simulation, then no
hysteresis will be observed. The second reason is that double
peaks in the probability density function have also been observed
near continuous transitions in finite systems, for examples in
4-states potts model in two dimensions. So the first three
methods, although  efficient for the case of strongly first order
transitions, are nor suitable to investigate the weakly first
order transitions. Methods (4) and (5) are the results of the
discontinuity  of internal energy  at first-order phase
transitions. Since the specific heat is obtained by derivative of
internal energy respect to temperature, we expect that it present
a delta function sigularity at the transition point. This causes
the specific heat peak to diverge as $L^d$, while its half-width
narrows like $L^{-d}$. Consequently, for the specific heat peak
and transition temperature, we will have the following behaviours
at a first order phase transition:

\begin{equation}\label{1st-c}
C_{max}(L)=c_{1}+c_{2}L^d
\end{equation}

\begin{equation}\label{1st-t}
T_{c}(L)=T_{c}(\infty)+AL^{-d}.
\end{equation}

The coefficient $c_{2}$ in eq.(\ref{1st-c}) is related to latent
heat per site through the following relation:

\begin{equation}\label{latent-c}
c_{2}=\frac{(e_{1}-e_{2})^2}{4T_{c}^2},
\end{equation}

where $e_{1}$ and $e_{2}$ are the values of energy per site at
the transition point a first order phase transition.  For a
continuous phase transition, where the correlation length grows
as $\xi\sim |T-T_{c}|^{-\nu}$ near a critical point,  the
behaviours of these two quantities are as :

\begin{equation}\label{2nd-c}
C_{max}(L)=c_{1}+c_{2}L^{\frac{\alpha}{\nu}}
\end{equation}

\begin{equation}\label{2nd-t}
T_{c}(L)=T_{c}(\infty)+AL^{-\frac{1}{\nu}},
\end{equation}

in which $\alpha$ is specific heat singularity exponent.

Method (6) is a test for the Gaussian nature of the probability
density function $P(E)$ at $T_{c}$. For a continuous transition,
$P(E)$ is expected to be Gaussian at, as well as away from
$T_{c}$. For a first-order transition, $P(E)$ will be double
peaked in infinite lattice size limit, hence deviation from being
Gaussian cause the minimum of $U(L)$ tends  $U^{*}$ to be less
than $2/3$ as $L\rightarrow\infty$. $U^{*}$ is related indirectly
to the latent heat. This is like the  method (3) but  much more
sensitive, in a sense that small splitting in $P(E)$ for the
infinite system that do not result in a double peak for small
lattices can be detected. Another advantage of this technique is
that the minimum of $U_{L}$ is expected to approach $2/3$ or
$U^{*}$ as power law in $L$, thus allowing one to extrapolate to
$L=\infty$ as:

\begin{eqnarray}\label{bind}
U_{4}(L)|_{min}=&&\frac{2}{3}-\left(e_{1}/e_{2}-e_{2}/e_{1}\right)^{2}/12+BL^{-d}+\\\nonumber
 &&O(L^{-2d}),
\end{eqnarray}

The eq.(\ref{bind}) implies that:

\begin{equation}\label{vstar}
U^{*}=\frac{2}{3}-\left(e_{1}/e_{2}-e_{2}/e_{1}\right)^2/12.
\end{equation}

For weakly first order transitions where latent heat per site is
too small ($\Delta e=e_{1}-e_{2}<<e_{1}$), we can write

\begin{equation}\label{latent-bind}
U^{*}\approx\frac{2}{3}-(\Delta e/e)^{2}/3.
\end{equation}

As an example we have used multi-histogram method (At least ten
histograms were combined for each lattice size) to calculate the
temperature dependent of $U_{4}(L)$ for $J_{1}=1.0$ , $J_{2}=0.8$
and $J_{3}=3.0$ depicted in Fig. (3), in which two minima exists
for all values of linear lattice sizes ($L=6,7,8,9,10,12$). The
right or high temperature minima indicate the transition from
disorder to hexatic phase for which, we will show in what follows,
that $U^{*}=2/3$, indicating a second order phase transition. The
left or low temperature minima represent the transition from
hexatic to hexatic plus herringbone phase. For this transition,
however $U^{*}$ turns to be  less than $2/3$ (table.I) showing
that is a first order transition.

Since no hysteresis, discontinuities or double peaked $P(E)$ were
observed in our simulation, we proceed to determine the order of
the transition by scaling of the specific heat with lattice size
and the determination of $U^{*}$ which is the most reliable
method.

\section{results and discussion}

In our work, at least five histograms were combined for each
lattice size for different temperatures near $T_c$. For each
histogram, we performed $5\times 10^5$ MCS for equilibration and
$1\times 10^6$ for data collection, while 10 to 20 monte calro
sweeps were discarded between successive measurements for
decreasing the correlation between them. Because the energy
spectrum is continuous, the data list obtained from a simulation
is basically a histogram with one entry per energy value. In
order to use the histogram method efficiently, we divide the
energy range $E\leq 0$ into 20 000 and  200 000 bins and
reconstructed the histograms. The results of the two binning
agreed with each other within statistical errors. therefore we
chose 20 000 bines throughout our simulation. In all simulations
we fixed $J_{1}=1.0$ and $J_{3}=3.0$ and changed values of $J_{2}$
from 0.5 to 1.3.

Starting from $J_{2}=0.5$, for all lattice sizes, we observed two
peaks in specific heat and two minima in the Binder forth energy
cumulant vs temperatures in cooling run ( see Fig.(1) and
Fig(3)). By increasing the value of $J_{2}$ those two peaks and
minima get closer to each other as for $J_{2}=0.8$ the first peak
change to be like a shoulder, while the two minima continue to be
well separated. This behaviour can be traced until $J_{2}=0.9$ for
which the two transitions merge to each other. For $J_{2}\geq
0.9$, also one peak and a minimum is obtained suggesting  that
$J_{2}=0.9$ can be considered as a critical end point in our
simulation, above which only one transition from disordered to
hexatic+herringbone phase would occur.

In what follows, we discuss separately the three transitions:
1)Isotropic-hexatic, 2)hexatic-hexatic+herringbone(locked phase)
3)Isotropic-hexatic+herringbone.

\subsection{Isotropic-hexatic transition }
Using the Binder forth energy cumulant to determine  the order of
transition, we found that for all of those transitions for
$J_{2}=0.5,0.6,0.7,0.8,0.85$,  the minimum value of $U_{L}(U^{*})$
tends to $2/3$ within the statistical error of the simulation.
For example in Fig.(4-a) we have ploted $U_{L}$ vs $L^{-3}$ for
$J_{2}=0.7$. The best fitting of the data to eq.(\ref{bind}), by
using least square procedure, shows that $U^{*}=0.66647(31)$
which is equal to $2/3$ within on e.s.d. This is true for all
isotropic-hexatic transition points (table I). These results show
that to the resolution of our simulation all of these transitions
are second order.

To calculate the critical exponents we used the scaling relation
of  the maximum values of heat capacity per site ($C_{max}$)
versus lattice sizes. The small range of the values of $C_{max}$
(i.e 2.54 for $L=6$ to 3.0 for $L=12$ for $J_{2}=0.7$ ) measured
for all  points along  this critical line, is the characteristic
of the transitions with cusp singularity in specific heat with
$\alpha\sim 0$. Figure(5-a) shows the best fit to $C_{max}$ as
power law in lattice size (eq.(\ref{2nd-c}), representing
$\alpha/\nu=-0.17(15)$ with relatively large error. However, The
calculating of the exact values of the critical exponent is not
our main purpose,  What is important for us is this point that
this transition line show no new universality class other than XY
universality.

For calculation of the critical temperatures, we used the power
law relation (\ref{2nd-t}) for fitting the effective transition
temperatures achieved by determining the location of specific heat
maxima and Binder cumulant minima (Fig(6-a). All the calculated
quantities discussed above, for this transition line, is listed
in Table.I.

\begin{table}[t]
\begin{tabular}[t]{|c|c|c|c|} 
$J_{2}$ &   $U^{*}$ & $T_{c}$ & $\alpha/\nu$ \\
\hline
 0.5    &  0.66656(34) & 2.16(4)  & -0.15(13)   \\
 0.6    &  0.66648(20) & 2.17(3)  & -0.13(10)   \\
 0.7    &  0.66648(31) & 2.16(5)  & -0.17(15)    \\
 0.8    &  0.66653(15) & 2.13(4)  & -0.13(12)    \\
 0.85   &  0.66655(30) & 2.16(4)  & --------   \\
 1.1    &  0.66660(8)  & 2.46(3)  & -0.10(3)  \\
 1.2    &  0.66664(15) & 2.53(6)  & -0.10(7)  \\
 1.3    &  0.66665(10) & 2.60(4)  & -0.11(9)  \\
\end{tabular}
\narrowtext\caption{Second order transitions. Calculated values
$U^{*}$ are obtained from fitting to eq.(\ref{bind}), $T_{c}$ from
eq.(\ref{1st-t}) and $\alpha/\nu$ from eq.(\ref{2nd-c}).}
\end{table}
\begin{table}[t]
\begin{tabular}[t]{|c|c|c|c|c|} 
$J_{2}$ &  $c_{2}$ & $U^{*}$ & $T_{c}$ & $\Delta e$ \\
\hline
 0.5    &  0.00353(35)    &  0.66630(10) & 1.254(3)  &  0.159(16)   \\
 0.7    &  0.00332(42)    &  0.66587(11) & 1.705(1)  &  0.209(24)   \\
 0.8    &  0.00230(8)     &  0.6660(30)  & 1.930(9)  &  0.185(37)   \\
 0.9    &  0.00225(28)    &  0.66558(31) & 2.110(8)  &  0.210(19)   \\
 0.95   &  0.00264(90)    &  0.66574(35) & 2.186(4)  &  0.213(39)   \\
 1.0    &  0.00267(50)    &  0.66476(10) & 2.283(7)  &  0.252(29)   \\
\end{tabular}
\narrowtext\caption{First order transitions. Calculated values of
straight line slope $c_{2}$ are obtained from fitting the data to
eq.(\ref{1st-c}), $U^{*}$ from eq.(\ref{bind}), $T_{c}$ from
eq.(\ref{1st-t}) and discontinuity of energy per site ($\Delta
e$) from averaging between eqs.(\ref{latent-c}) and
(\ref{latent-bind}).}
\end{table}

\subsection{hexatic to hexatic+herringbone transition }
The transition from hexatic phase with long range XY order to
hexatic+herringbone phase, which possess the three state potts
symmetry, is known to be a in the 3-state potts universality
class in 3D and hence weakly first order. This is verified by the
procedure discussed in previous subsection. Figures (4-b),(5-b)
and (6-b) show the size dependence of $U_{L}$, $C_{max}$ and
$T_{c}$ for $J_{2}=0.7$. As it can be seen $U^{*}=0.66578(10)$
which is less than $2/3$ within one e.s.d. The latent heat per
site averaged from eqs.(\ref{latent-c}) and (\ref{latent-bind})
is derived to be about 0.21 in the units of $J_{1}$. The
calculated quantities for other values of $J_{2}$ (0.5,0.8,0.90)
has been listed in table.II. In the resolution of our simulation,
$J_{2}=0.9$ is the end point of the isotropic-hexatic critical
line.

\subsection{isotropic to hexatic+herringbone transition }
For $J_{2}>0.9$ only one transition would appear, in which the
hexatic and herringbone orders establish simultaneously. It can
be seen from the data listed in tables I and II that this
transition is first order for $J_{2}=0.9,0.95,1.0$, while it
changes to second order for $J_{2}=1.1,1.2,1.3$.  The size
dependence of $U_{L}$, $C_{max}$ and $T_{c}$ for $J=2=1.0$ and
$J2=1.2$, together with the best fits on the data, have been shown
in figures (7) to (12). As it is seen from the table.I, all
specific heat exponents calculated for $J_{2}>1.1$ are negative
and equal up to the measurement errors, suggesting all belong the
the same universality class.The other important result here is
the existence of a tricritical point located  between $J_{2}=1.0$
and $J_{2}=1.1$. In figure.13 the phase-diagram of the BA
Hamiltonian, obtained from Monte Carlo simulation has been
depicted.

\section{Conclusion }

In summary, employing the optimized Monte Carlo simulation based
on multi-histogram method, we investigated the phase diagram
associated with the Hamiltonian purposed by Bruinsma and Aeppli,
which  consists of two coupled XY order parameters (indicating
hexatic and short range herringbone orders), in the regime that
the two order parameters are coupled  strongly. The simulation
reveals three distinct phases for this model. According to the
simulation results, the transition from isotropic to only haxatic
phase remains second order all over on this transition line,
ruling out the existence of any tricritical point on this line.
It is also found that the transition from hexatic to locked phase
(hexatic+herringbone) is always weakly first order. These two
transition lines meet each other at a critical end point
characterizing by ${J_{2}\over J_{1}}=0.9$ and ${T_{c} \over
J_{1}} =2.110(8)$. For ${J_{2}\over J_{1}}>0.9$ however, only one
transition occurs from isotropic to locked phase whose order
found to be weakly first order up to ${J_{2}\over J_{1}}=1.0$ and
turned  to be second order for ${J_{2}\over J_{1}}\geq 1.1$, for
which  all calculated specific-heat exponents are negative and
equal within the simulation errors. It shows that all these
continues transitions are in the same universality class.However,
for the interval $1.0<{J_{2}\over J_{1}}<1.1$, there may be the
possibility that the heat capacity critical exponent ($\alpha)$
exhibits an evolution from being negative for ${J_{2}\over
J_{1}}=1.1$ to a large positive value near ${J_{2}\over
J_{1}}=1.0$. Checking this idea requires  more accurate  and
higher resolution  simulations to determine the critical exponents
and is the subject of our present research.

The last result then also suggests the existence of a tricritical
point in between ${J_{2}\over J_{1}}=1.0$ and ${J_{2}\over
J_{1}}=1.1$, providing  a plausible explanation for large heat
capacity anomaly exponents, observed in the experiments, in terms
of occurrence of  SmA-HexB transition(which in our simulation is
represented as transition from the disorder phase to a phase
consists of both long range hexatic and short range herringbone
orders), near this  tricritical point. Knowing that $d=3$ is the
upper critical dimension for tricritical point, The deviation of
experimentally measured heat capacity exponent ($\alpha\sim 0.6$)
from mean-field  value $\alpha=0.5$ may be related to the
logarithmic corrections arising from marginal fluctuations at the
tricritical point. However, While it is a convincing argument,
this question remains that why seven different liquid crystal
compounds nmOBC and five binary mixtures n(10)OBC, with very
different SmA-HexB temperature ranges(which effect the coupling
of two order types) yield approximately the same value
$\alpha\approx0.6$ and should all be in the immediate vicinity of
a particular thermodynamic point.

As an open problem, we  address the study of  weak coupling model
which might be important for the case of  SmA-HexB transition in
the mixture of 3(10)OBC and PHOAB  that possess a very large
temperature range for the HexB phase above the crystallization
temerature to the CryE phase, Yet exhibits the same unusual
critical exponents{\cite{huang}}.

Another important issue is  the possibility of the existence of
long-range herringbone order in a system with long-range
orientational order and short-range translational order, as
suggested by thin-film heat capacity data {\cite{huang}}.

We finally hope that our work will motivate further theoretical,
numerical and experimental investigations of this very
interesting problem.

{\bf Acknowledgment} We would like to thank M.J.P Gingras for
very useful comments and discussions. F.Shahbazi was financially
supported in part by IUT grant No-1PHB821.


\newpage
\begin{figure}
\epsfxsize=8.5truecm \epsfbox{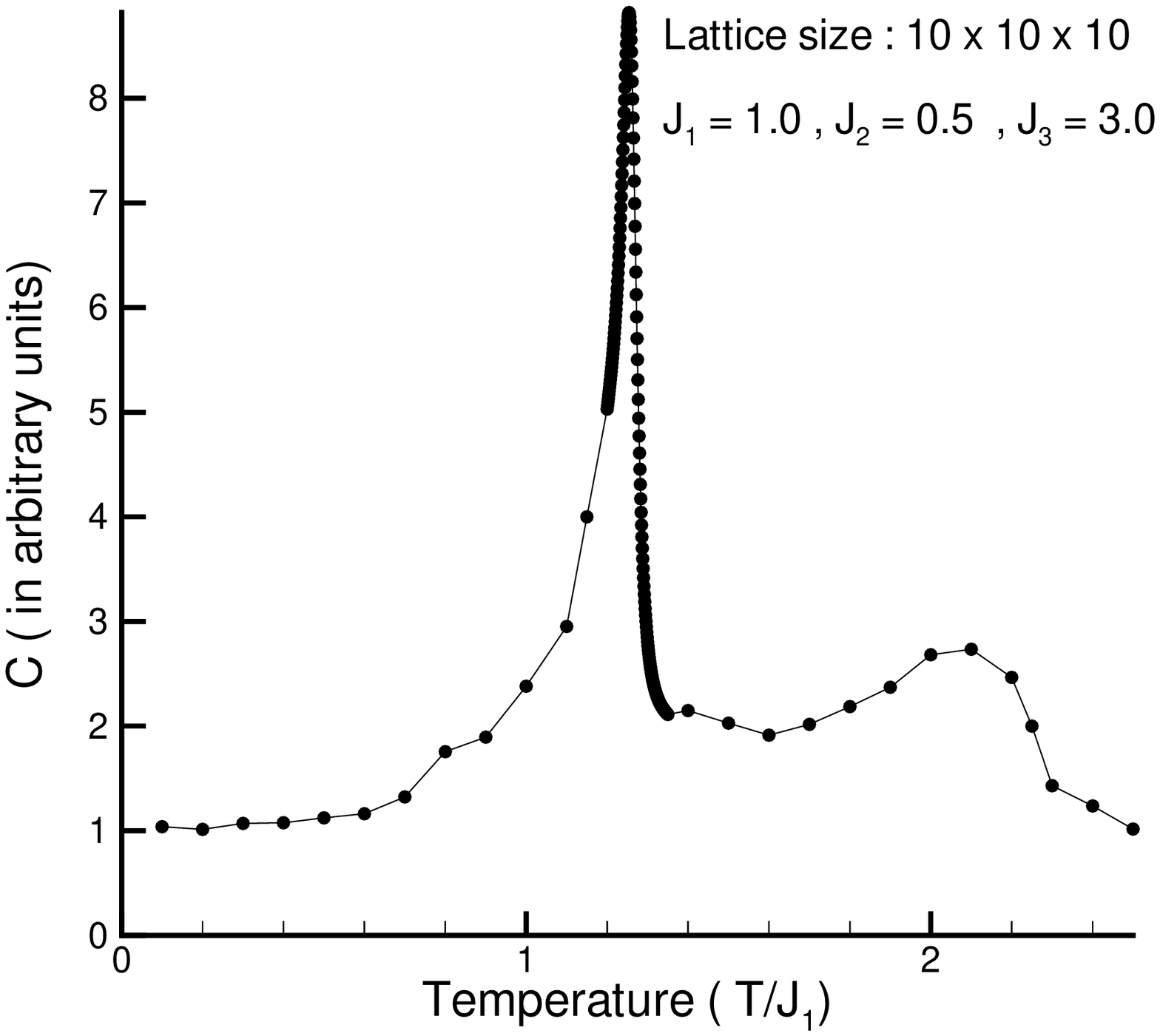} \narrowtext \caption{
Temperature dependence of specific heat for $J_{1}=1.0$,
$J_{2}=0.5$ and $J_{3}=3.0$. The points between $T=1.35$ and
$T=1.2$ has been derived using multi-histogram method(see the
text). }
\end{figure}

\begin{figure}
\epsfxsize=8.5truecm \epsfbox{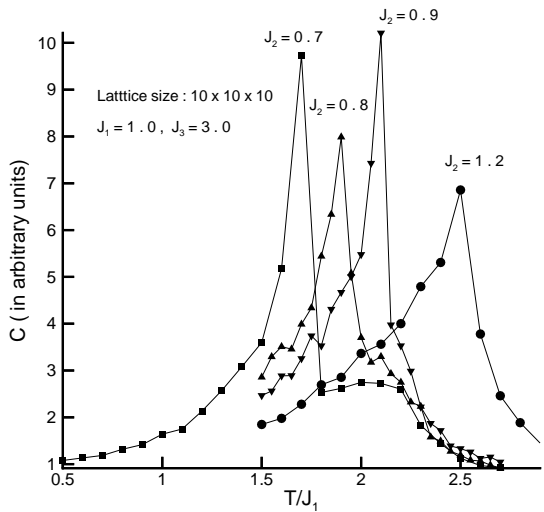} \narrowtext \caption{
Temperature dependence of specific heat for $J_{1}=1.0$,
$J_{3}=3.0$ and $J_{2}=0.7,0.8,0.9,1.2$ }
\end{figure}
\newpage
\begin{figure}[t]
\epsfxsize=8.5truecm \epsfbox{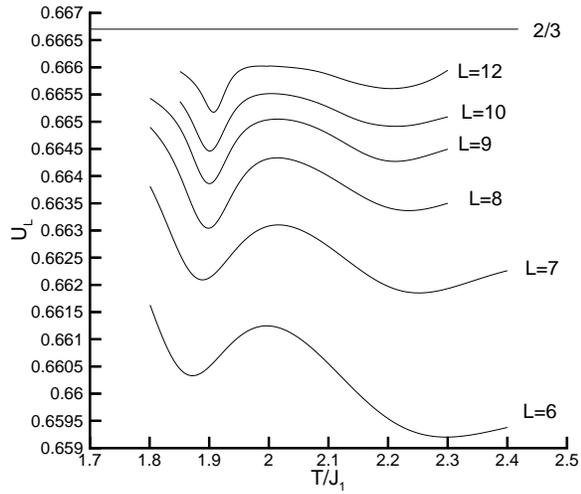} \narrowtext \caption{
Binder's fourth energy cumulant for $J_{1}=1.0$, $J_{2}=0.8$ and
$J_{3}=3.0$. High temperature minima are near the  transition from
isotropic to hexatic while the low temperature minima indicate
the  transition from hexatic to hexatic+herringbone state. }
\end{figure}

\newpage
\begin{figure}[t]
\epsfxsize=7.truecm \epsfbox{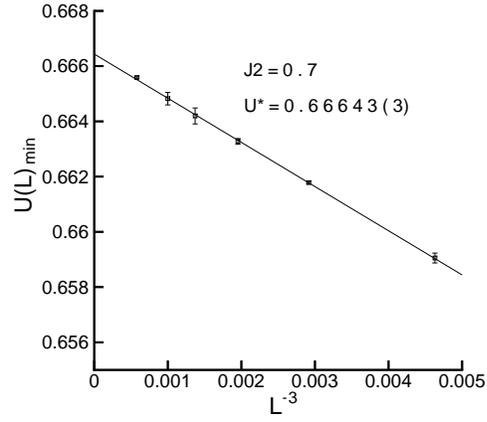}
\epsfxsize=7.truecm\epsfbox{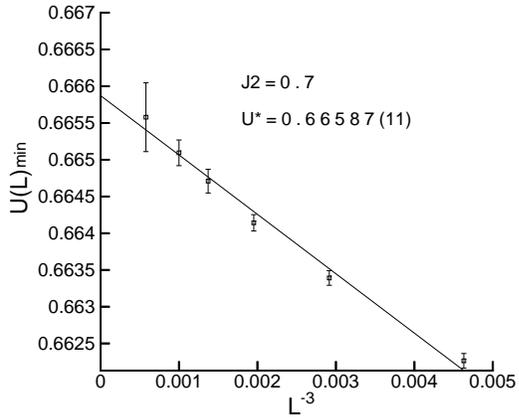} \narrowtext
\caption{Size dependence of binder fourth energy cumulant minima,
 calculated by optimized re-weighting for $J_{1}=1.0$,
$J_{2}=0.7$ and $J_{3}=3.0$.(a)Transition from isotropic to
hexatic phase (second order),(b) transition from hexatic to
hexatic+herringbone (first order). Solid lines represent fits to
(\ref{bind}).}
\end{figure}

\newpage

\begin{figure}[t]
\epsfxsize=7.truecm \epsfbox{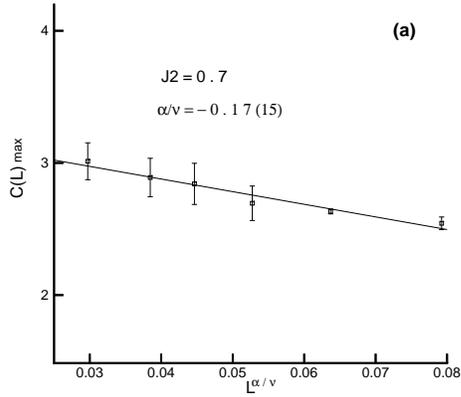}
\epsfxsize=7.truecm\epsfbox{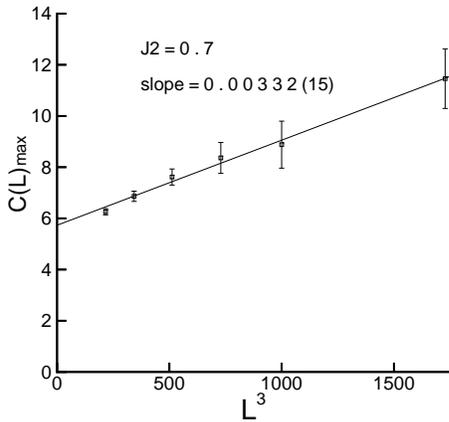}
 \narrowtext \caption{
Size dependence of the specific heat maxima, $C_{max}$, calculated
by optimized re-weighting for $J_{1}=1.0$, $J_{2}=0.7$ and
$J_{3}=3.0$.(a)Transition from isotropic to hexatic phase,(b)
transition from hexatic to hexatic+herringbone phases. Solid lines
represent fits to (\ref{2nd-c}) for (a) and (\ref{1st-c}) for (b)
. }
\end{figure}
\begin{figure}[t]
\epsfxsize=7.truecm \epsfbox{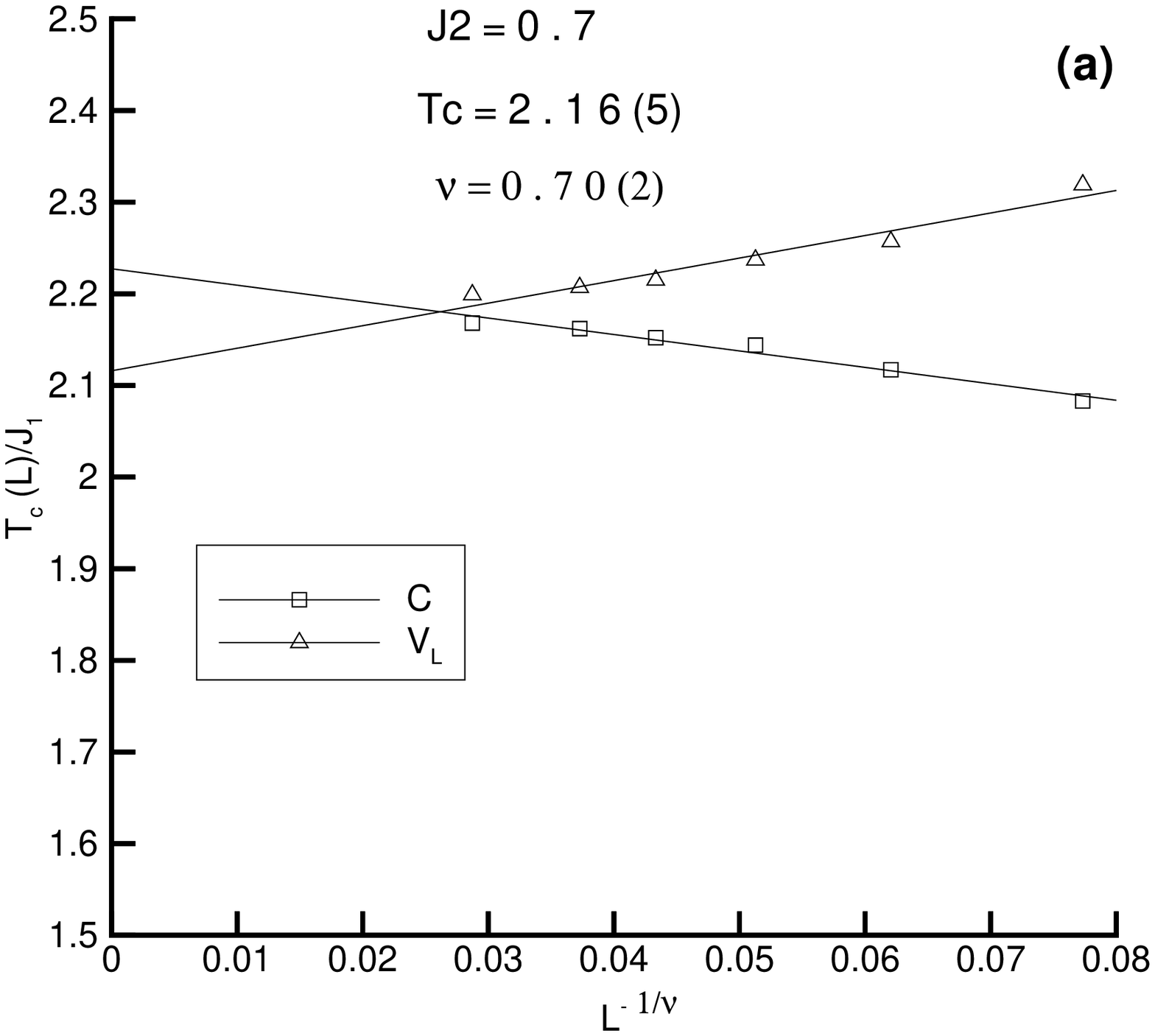}
\epsfxsize=7.truecm\epsfbox{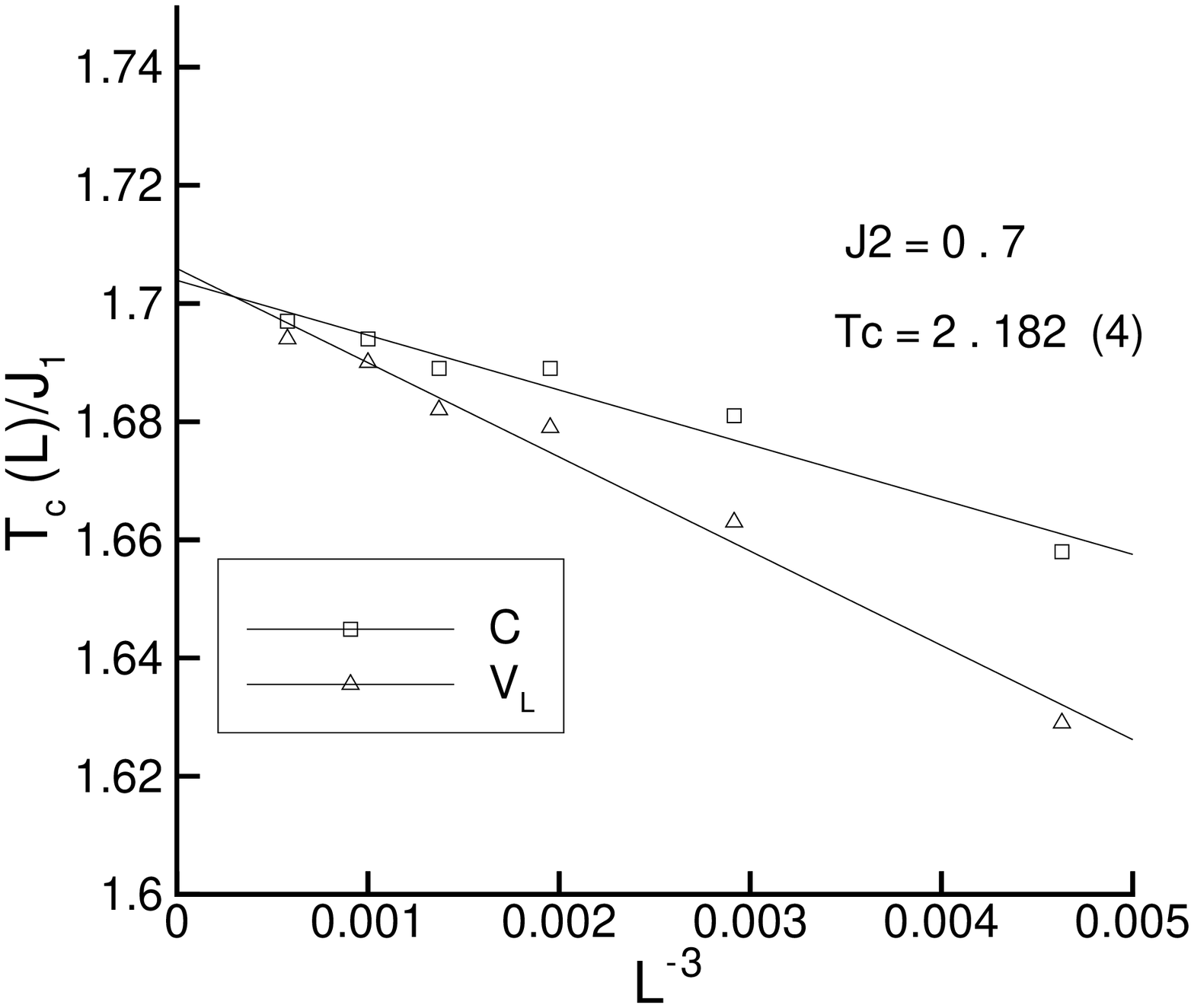}
 \narrowtext \caption{
Scaling  of the effective transition temperatures with lattice
size. for $J_{1}=1.0$, $J_{2}=0.7$ and $J_{3}=3.0$. The $T_{c}$'s
were obtained from the location of the maxima of specific heat and
minima of Binder fourth energy cumulant. (a)Transition from
isotropic to hexatic phase ,(b) transition from hexatic to
hexatic+herringbone phases. The solid lines represent fits to
(\ref{2nd-t}) for (a) and (\ref{1st-t}) for (b). }
\end{figure}

\newpage

\begin{figure}[t]
\epsfxsize=7.truecm \epsfbox{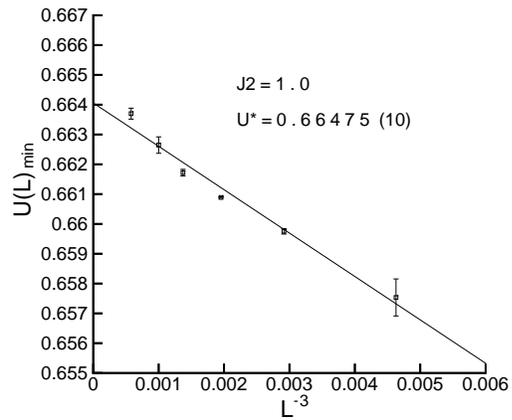} \caption{Size
dependence of binder fourth energy cumulant minima, calculated by
optimized re-weighting for $J_{1}=1.0$, $J_{2}=1.0$ and
$J_{3}=3.0$ at the transition point from isotropic to
hexatic+herringbone phases. Solid line represent fit to
(\ref{bind}) the obtained value $U^{*}=0.66475(10)<2/3$  indicates
a first order transition.}
\end{figure}

\newpage

\begin{figure}[t]
\epsfxsize=7.truecm \epsfbox{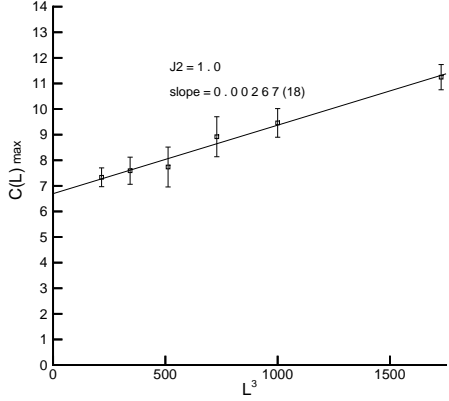}
 \narrowtext \caption{
Size dependence of the specific heat maxima, $C_{max}$, calculated
by optimized re-weighting for $J_{1}=1.0$, $J_{2}=1.0$ and
$J_{3}=3.0$ at the transition point from isotropic to
hexatic+herringbine phases. Solid line represents fit to
(\ref{1st-c}). }
\end{figure}

\begin{figure}[t]
\epsfxsize=7.truecm \epsfbox{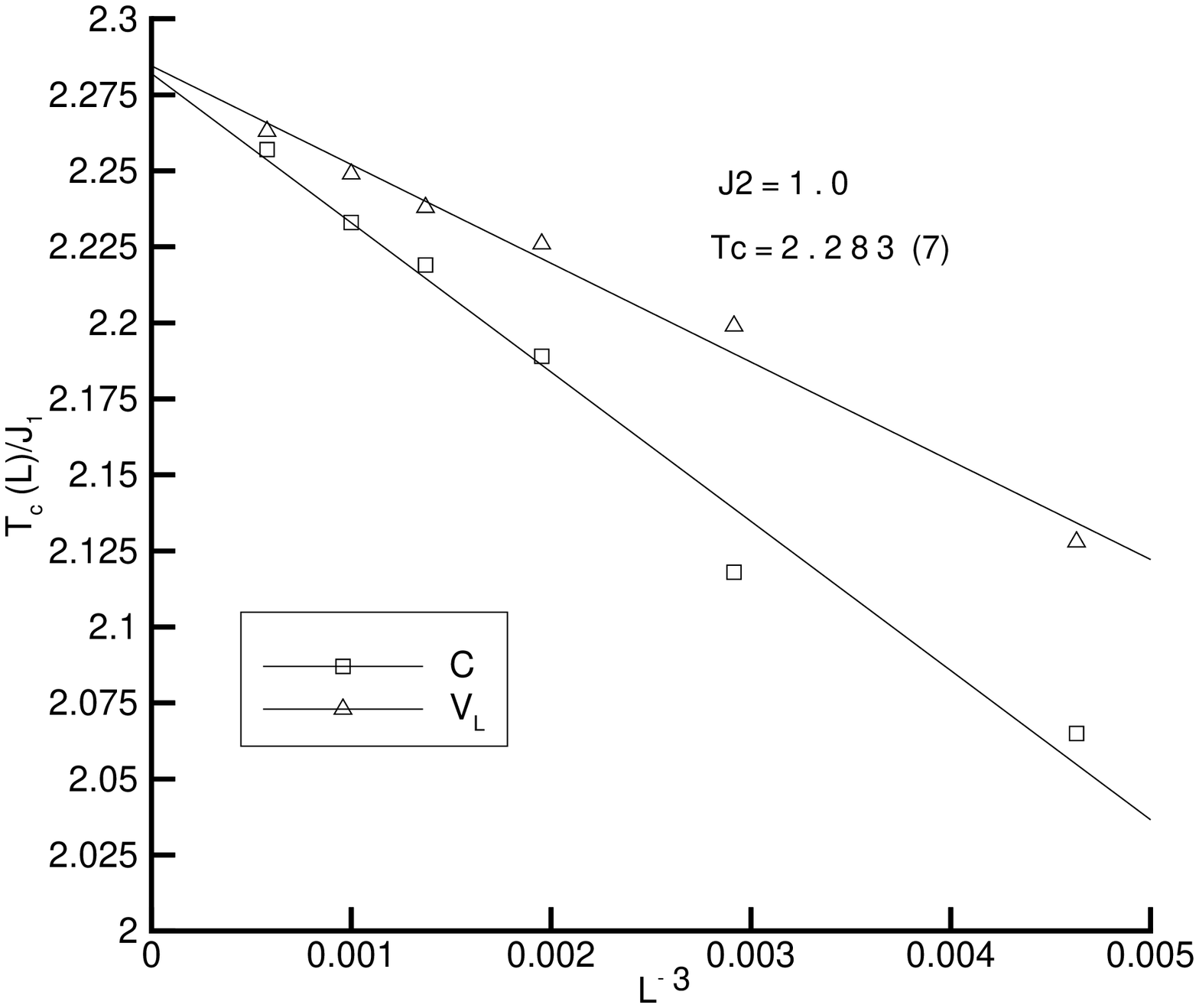}
 \narrowtext \caption{
Scaling  of the effective transition temperatures with lattice
size, for $J_{1}=1.0$, $J_{2}=1.0$ and $J_{3}=3.0$. The $T_{c}$'s
were obtained from the location of the maxima of specific heats
and minima of Binder fourth energy cumulants. Solid lines
represent fit (\ref{1st-t}).}
\end{figure}

\newpage
\begin{figure}[t]
\epsfxsize=7.truecm \epsfbox{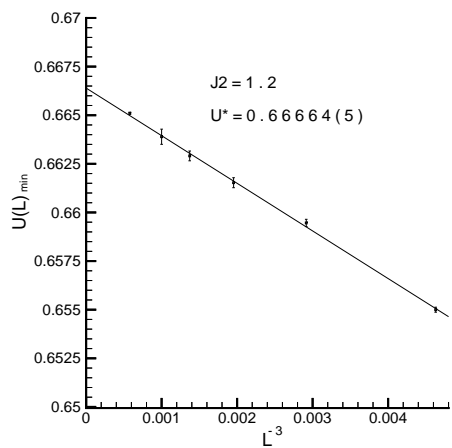} \caption{Size
dependence of binder fourth energy cumulant minima, calculated by
optimized reweighting for $J_{1}=1.0$, $J_{2}=1.2$ and
$J_{3}=3.0$. Solid line represent fit to (\ref{bind}).}
\end{figure}

\newpage

\begin{figure}[t]
\epsfxsize=7.truecm \epsfbox{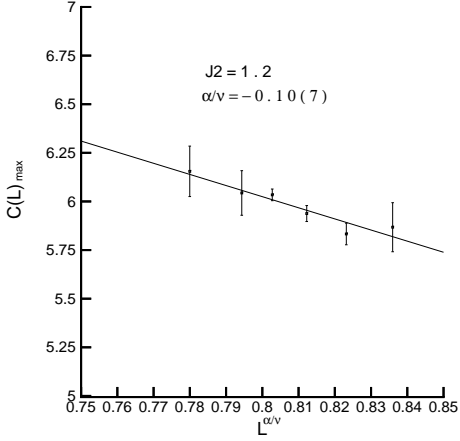}
 \narrowtext \caption{
Size dependence of the specific heat maxima, $C_{max}$, calculated
by optimized re-weighting for $J_{1}=1.0$, $J_{2}=1.2$ and
$J_{3}=3.0$ at the transition point from isotropic to
hexatic+herringbine phase. Solid line represents fit to
(\ref{2nd-c}), indicating a second order transition with negative
value for specific heat anomaly exponent $\alpha$. }
\end{figure}

\begin{figure}[t]
\epsfxsize=7.truecm \epsfbox{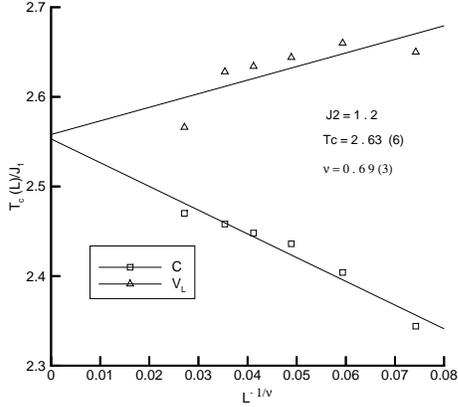}
 \narrowtext \caption{
Scaling  of the effective transition temperatures with lattice
size. for $J_{1}=1.0$, $J_{2}=1.2$ and $J_{3}=3.0$. The $T_{c}$'s
were obtained from the location of the maxima of specific heats
and minima of Binder fourth energy cumulants. Solid lines
represent fit (\ref{2nd-t}) with value $0.69(3)$ for exponent
$\nu$.}
\end{figure}

\begin{figure}[t]
\epsfxsize=8.5truecm \epsfbox{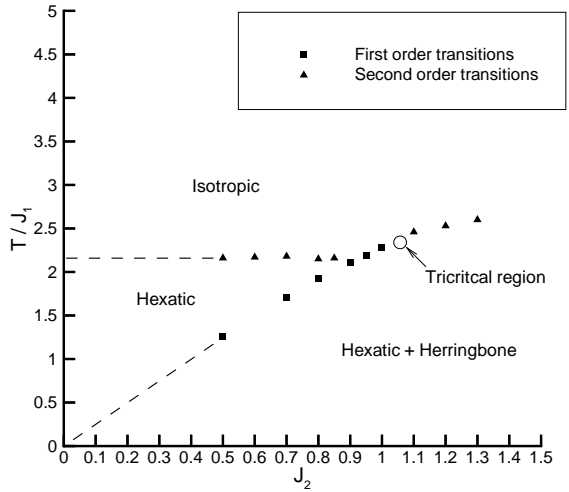} \narrowtext
\caption{ Schematic of the phase diagram obtained from
simulation. Transition temperatures (in units of $J_{1})$ versus
$J_{2}$. Three phases Isotropic, Hexatic and Hexatic+Herringbone
has been shown and the dashed lines are just  representing the
separating of distinct phases. The region specified by circle is
the tricritical region where order of the transition changes from
being first order for $J_{2}=1.0$ to second order for $J_{2}=1.1$
}
\end{figure}

\end{document}